\documentclass[prb, reprint, amsfonts, amssymb, amsmath, superscriptaddress]{revtex4-1}
\usepackage[english]{babel}
\usepackage[utf8]{inputenc}
\usepackage{xcolor}
\usepackage{graphicx}
\usepackage{multirow}
\usepackage{xspace}
\usepackage{hyperref}
\hypersetup{colorlinks=true, linkcolor=black, citecolor=green!50!black, urlcolor=blue!50!black}

\newcommand{\CrAlC}{Cr$_2$AlC\xspace}
\newcommand{\PtCoO}{PtCoO$_2$\xspace}
\newcommand{\YCoB}{YCo$_3$B$_2$\xspace}

\let\hatOrig\hat
\renewcommand{\vec}[1]{\boldsymbol{\mathbf{#1}}}
\renewcommand{\hat}[1]{\boldsymbol{\mathbf{\hatOrig{#1}}}}
\renewcommand{\Im}{\operatorname{Im}}
\renewcommand{\Re}{\operatorname{Re}}
\newcommand{\D}{\mathrm{d}}
\newcommand{\sub}[1]{\ensuremath{_{\textrm{#1}}}} \newcommand{\super}[1]{\ensuremath{^{\textrm{#1}}}}  

\newcommand{\affilMSE}{\affiliation{Department of Materials Science \& Engineering, 110 8$^{th}$ St, Troy, NY 12180, USA}}
\newcommand{\affilPhy}{\affiliation{Department of Physics, Applied Physics, and Astronomy, 110 8$^{th}$ St, Troy, NY 12180, USA.}}

\begin{document}
\title{Fermi surface anisotropy in plasmonic metals increases\\
the potential for efficient hot carrier extraction}

\author{Sushant Kumar}
\affilMSE

\author{Christian Multunas}
\affilPhy

\author{Ravishankar Sundararaman}
\email{sundar@rpi.edu}
\affilMSE
\affilPhy

\date{\today}

\begin{abstract}
Realizing the potential of plasmonic hot carrier harvesting for energy conversion and photodetection requires new materials that resolve the bottleneck of extracting carriers prior to energy relaxation within the metal.
Using first-principles calculations of optical response and carrier transport properties, we show that directional conductors with Fermi velocities restricted predominantly to one or two directions present significant advantages for efficient hot carrier harvesting.
We show that the optical response of film-like conductors, \PtCoO and \CrAlC, resemble that of 2D metals, while that of wire-like conductors, CoSn and \YCoB, resemble that of 1D metals, which can lead to high mode confinement and efficient light collection in small dimensions, while still working with 3D materials with high carrier densities.
Carrier lifetimes and transport distances in these materials, especially in \PtCoO and CoSn, are competitive with noble metals.
Most importantly, we predict that carrier injection efficiency from all of these materials into semiconductors can exceed 10\% due to the small component of carrier momentum parallel to the metal surface, substantially improving upon the typical less than 0.1\% injection efficiency from noble metals into semiconductors.
\end{abstract}


\maketitle
\section{Introduction}  \label{sec:intro}
High-energy electrons and holes excited in metal nanostructures upon the decay of plasmons provide a pathway to leverage light collected efficiently at the nano scale for photochemistry,\cite{HotCarrierSolarChemical} imaging,\cite{HotCarrierImaging} and photovoltaic energy conversion.\cite{HotCarrierSolarCell}
A continuing challenge in plasmonic hot carrier applications is utilizing the excited carriers prior to energy relaxation in the metal,\cite{HotCarrierReview1, HotCarrierReview2} requiring nanostructures smaller than the mean free paths of carriers and material interfaces that are capable of extracting a large fraction of carriers from the metal.\cite{ScratchTheSurface}

The field of plasmonic hot carriers has made great strides in expanding the space of materials from noble metals to include refractory plasmonic materials\cite{RefractoryPlasmonics} such as transition metal nitrides,\cite{PlasmonicNitridesReview, NitrideHotElectron, NitrideCarriers} and low-dimensional materials such as MXenes.\cite{MXenesForPlasmonic2018, MXenesPhotoDetect2019, MXeneMoS2HeteroPhotoDetect2019}
These materials remain stable at smaller dimensions than noble metals, making it possible to minimize energy losses during the transport of carriers across the nanostructure.\cite{NitrideCarriers}
Additionally, lower-dimensional plasmonic materials can exhibit higher mode confinement than 3D metals, allowing efficient light capture at smaller length scales.\cite{Argentene}
However, the overall efficiency of extracting hot carriers from plasmonic metals into semiconductors for photovoltaic and photodetection applications remains low, primarily due to the mismatch between carrier momenta near the Fermi energy in metals and the band edges in semiconductors,\cite{HotCarrierReflection, GoldGaNcarriers, CuGaNcarriers} requiring further advances in materials for plasmonic hot carriers.\cite{ScratchTheSurface}

Here, we propose that directional conductors -- 3D materials that behave like lower-dimensional metals due to Fermi velocities restricted to one or two dimensions -- have the potential to simultaneously absorb light efficiently at nanoscale dimensions and inject carriers with high probability into semiconductors.
These directional conductors typically have highly anisotropic crystal structures (e.g., layered-materials) which lead to highly anisotropic Fermi surfaces with directional properties.
Using first-principles calculations of electronic structure, optical response, electron-phonon and electron-electron scattering, we show that the best directional conductors, including \PtCoO and CoSn, are competitive with noble metals in their frequency range of plasmonic response and carrier transport distances.
We show that their optical response resembles lower-dimensional metals, which may allow the design of plasmonic nanostructures with more efficient light collection at small dimensions than their 3D counterparts.
Finally, we show that the lower magnitude of Fermi momenta and the strong directional distribution can be leveraged to greatly enhance the probability of carrier injection into the semiconductor: from below 0.1\% typical for noble metals to over 10\% for the directional conductors as predicted by a modified Fowler model.
 
\section{Methods}

\subsection{Computational details}
\label{sec:MethodsDetails}

We perform electronic structure calculations using the open-source JDFTx software for density-functional theory in the plane wave basis.\cite{JDFTx}
We use the Perdew-Burke-Ernzerhof generalized gradient approximation to the exchange correlation functional,\cite{PBE} with non-relativistic ultrasoft Garrity-Bennett-Rabe-Vanderbilt (GBRV) pseudopotentials at kinetic energy cutoffs of 20 Hartrees for the wavefunctions and 100 Hartrees for the charge density.\cite{GBRV}
We optimize lattice parameters and internal geometries self-consistently for each material, and calculate phonons and electron-phonon coupling from first-principles.
The converged lattice parameters have been compared with the corresponding experimental measurements in Table~\ref{tab:params}.
We use a $k$-mesh of   12$\times$12$\times$2 for \PtCoO and 12$\times$12$\times$12 for \CrAlC, CoSn, and \YCoB.
For our phonon calculations, we use a $q$-point sampling of 3$\times$3$\times$1 for \PtCoO, 3$\times$3$\times$2 for \CrAlC and 3$\times$3$\times$3 for CoSn and \YCoB.

\begin{table}
\centering
\setlength{\tabcolsep}{3pt}
\begin{tabular}{cccccc}
\hline
Material & $N\sub{atoms}$ & $a$~[\AA] & $c$~[\AA]  \\
\hline
\PtCoO & 12 & 2.85 (2.82) & 18.03 (17.81)\\
\CrAlC & 8 & 2.85 (2.86) & 12.75 (12.82) \\
CoSn   & 6 & 5.26 (5.28)& 4.26 (4.26) \\
\YCoB  & 6 & 5.00 (5.04) & 3.04 (3.03) \\
\hline
\end{tabular}
\caption{Number of atoms per unit cell ($N\sub{atoms}$) and converged lattice parameters for each material. The experimental lattice parameters ($a$ and $c$) for \PtCoO~\cite{kushwaha2015nearly}, \CrAlC~\cite{lin2005situ}, CoSn~\cite{meier2020flat} and \YCoB~\cite{kowalczyk2000transport} have been provided for comparison next to the DFT-predicted values (in parenthesis).}
\label{tab:params}
\end{table}

We construct maximally localized Wannier functions for each material to accurately reproduce the electronic band structure up to at least 40~eV above the Fermi level, and use them to transform electron and phonon properties using the Wannier module of JDFTx.\cite{MLWFmetal,JDFTx}
For each of these materials, we use a combination of atomic orbitals and Gaussian orbitals ($\sigma = 1 $ Bohr) as the initial guesses or trial orbitals.
The Gaussian orbitals are centered at random positions inside the unit cell.
We use 5 $d$-orbitals per Pt and Co atom, 1 $s$-orbital and 3 $p$-orbitals per O atom,  and 105 Gaussian orbitals for \PtCoO (159 in total).
For \CrAlC, we use 103 Gaussian orbitals.
We use 5 $d$-orbitals per Co and 5 $d$-orbitals and 3 $p$-orbitals per Sn atom, and 100 Gaussian orbitals for the case of CoSn (139 in total). Lastly, for \YCoB we use 5 $d$-orbitals per Co, 5 $d$-orbitals and 3 $p$-orbitals per Y atom, 1 $s$-orbital and 3 $p$-orbitals per B atom, and 90 Gaussian orbitals (121 in total).
Fig.~S3 in the supplemental information\cite{SI} shows that the real-space representations of the electronic Hamiltonian, dynamical matrix and electron-phonon matrix elements all decay rapidly with distance,\cite{ponce2021first} falling to $10^{-5} - 10^{-6}$ of their maximum values by the edge of the corresponding supercells.

Using these, we interpolate electronic energies $\varepsilon_{\vec{k}n}$ at wave vector $\vec{k}$ and band $n$, their band velocities $v_{\vec{k}n}$ and velocity matrix elements $v_{\vec{k}n'n}$, phonon frequencies $\omega_{\vec{q}\alpha}$ at wavevector $\vec{q}$ and polarization $\alpha$, and the electron-phonon matrix elements $g^{\vec{k}'-\vec{k},\alpha}_{\vec{k}'n',\vec{k}n}$ to significantly finer $\vec{k}$ and $\vec{q}$ meshes for integrals over the Brillouin Zone (BZ) in the calculation of optical and carrier transport properties described next.\cite{PhononAssisted}

\subsection{Dielectric functions}
\label{sec:MethodsDielectric}

We calculate the complex dielectric tensor,
\begin{equation}
\bar{\epsilon}(\omega) = 1 + \frac{4\pi i(\bar{\sigma}_0/\tau_{D0})}{\omega(\tau_D^{-1}(\omega) - i\omega)} + \bar{\epsilon}_d(\omega),
\label{eqn:epsilon}
\end{equation}
where $\bar{\sigma}_0$ and $\tau_{D0}$ are the frequency-independent (static) conductivity tensor and Drude lifetime respectively, $\tau_D$ is the frequency-dependent momentum relaxation time. The second term  captures the Drude response including the effect of phonon-assisted intraband transitions, while the final term $\bar{\epsilon}_d(\omega)$ captures the effect of direct optical transitions.
We evaluate the final term directly using Fermi's golden rule for the real part,\cite{DirectTransitions}
\begin{multline}
\Im\bar{\epsilon}_d(\omega)
= \frac{4\pi^2 e^2}{\omega^2} \int\sub{BZ} \frac{g\sub{s}\,\D\vec{k}}{(2\pi)^3} \sum_{n'n}
	(f_{\vec{k}n} - f_{\vec{k}n'})
\\ \times
	\delta(\varepsilon_{\vec{k}n'} - \varepsilon_{\vec{k}n} - \hbar\omega)
	\left(\vec{v}^{\ast}_{\vec{k}n'n} \otimes \vec{v}_{\vec{k}n'n}\right),
\label{eqn:sigmaDirect}
\end{multline}
where $g_s = 2$ is the spin degeneracy factor and $f_{\vec{k}n}$ are the Fermi occupations of each electronic state.
We then evaluate $\Re\bar{\epsilon}_d(\omega)$ from it using the Kramers-Kronig relation.
The energy-conserving $\delta$-function above is broadened to a Lorentzian due to electron-electron and electron-phonon scattering linewidths,
which we also calculate from first principles as described below in Section~\ref{sec:MethodsLifetime}.\cite{PhononAssisted}

In the Drude term above (second term of \eqref{eqn:epsilon}), we account for phonon-assisted transitions within the frequency-dependent momentum relaxation rate,
\begin{multline}
\tau_{\mathrm{D}}^{-1}(\omega) = \frac{2\pi}{\hbar g(\varepsilon_{\mathrm{F}}) b_T(\hbar\omega)}
	\sum_{\alpha\pm} \int_{BZ} \frac{\D\vec{q}}{(2\pi)^d}
\\ \times
		G^p_{\vec{q}\alpha} \frac{\pm b_T(\hbar\omega\pm\hbar\omega_{\vec{q}\alpha})}{e^{\pm\hbar\omega_{\vec{q}\alpha}/k_{\mathrm{B}}T} - 1},
\end{multline}
derived from the Eliashberg spectral function,\cite{AllenEliashberg} and generalized to non-zero temperature.\cite{Argentene}
Here, $g(\varepsilon_{\mathrm{F}})$ is the density of electronic states at the Fermi level, $b_T(\varepsilon) \equiv \varepsilon/(1-\mathrm{e}^{-\varepsilon/k_{\mathrm{B}}T})$ and
\begin{multline}
G^p_{\vec{q}\alpha} \equiv \sum_{nn'}\int_{BZ}\frac{g_s\Omega\,\D\vec{k}}{(2\pi)^d}
	\left| g^{\vec{q}\alpha}_{(\vec{k}+\vec{q})n',\vec{k}n} \right|^2
	\delta(\varepsilon_{\vec{k}n}-\varepsilon_{\mathrm{F}})
\\\times
\delta(\varepsilon_{(\vec{k}+\vec{q})n'}-\varepsilon_{\mathrm{F}})
    \left(1 - \frac{\vec{v}_{\vec{k}n}\cdot\vec{v}_{(\vec{k}+\vec{q})n'}}{|\vec{v}_{\vec{k}n}| |\vec{v}_{(\vec{k}+\vec{q})n'}|}\right),
\end{multline}
is the weight of each phonon mode in the `transport
Eliashberg spectral function'.
Above, $\Omega$ is the unit cell volume.
(The final factor dependent on velocities above accounts for scattering angle compared to the conventional Eliashberg spectral function.\cite{AllenEliashberg})
Finally, the numerator in the first term of \eqref{eqn:epsilon} is\cite{Argentene}
\begin{equation}
\frac{\bar{\sigma}_0}{\tau_{D0}} = \int_{BZ}\frac{e^2g_s\,\D\vec{k}}{(2\pi)^3} \sum_n
	\delta(\varepsilon_{\vec{k}n} - \varepsilon_{\mathrm{F}}) (\vec{v}_{\vec{k}n}\otimes\vec{v}_{\vec{k}n}),
\end{equation}
which is essentially the generalization of $e^2g(\varepsilon_{\mathrm{F}})v_{\mathrm{F}}^2/3$ from a spherical Fermi surface to the general anisotropic case.

We apply Monte Carlo sampling with a typical density of $\sim 10^7$ $\vec{k}$-values to fully converge the Brillouin zone integrals above, leveraging the Wannier interpolation of matrix elements.
Histogramming the integrands in the above expressions with respect to $\epsilon_{\vec{k}n}$, we also evaluate the energy distributions of carriers generated upon absorption.
See Ref.~\citenum{PhononAssisted} for further details and demonstration of the quantitative accuracy of this method for conventional plasmonic metals, as well as for low-dimensional and refractory plasmonic materials.\cite{GraphiteHotCarriers, NitrideCarriers}

\subsection{Carrier lifetimes and mean free paths}
\label{sec:MethodsLifetime}

The electron-phonon contribution to the inverse lifetime is given by Fermi's golden rule as
\begin{multline}
\left(\tau\sub{e-ph}^{-1}\right)_{\vec{k}{n}} =
\frac{2\pi}{\hbar} \int\sub{BZ} \frac{\Omega d\vec{k}'}{(2\pi)^3} \sum_{n'\alpha\pm}
	\delta(\varepsilon_{\vec{k}'n'} - \varepsilon_{\vec{k}n} \mp \hbar\omega_{\vec{k}'-\vec{k},\alpha})
\\
\times
	\left( n_{\vec{k}'-\vec{k},\alpha} + \frac{1}{2} \mp \left(\frac{1}{2} - f_{\vec{k}'n'}\right)\right)
	\left| g^{\vec{k}'-\vec{k},\alpha}_{\vec{k}'n',\vec{k}n} \right|^2,
\label{eqn:tauInv_ePh}
\end{multline}
where the factors in order correspond to energy conservation, occupation factors and 
the electron-phonon matrix elements.
The sum over $\pm$ counts phonon absorption and emission processes.
See Ref.~\citenum{PhononAssisted} and \citenum{GraphiteHotCarriers} for further details.

The electron-electron contribution is calculated using
\begin{multline}
\left(\tau_{e-e}^{-1}\right)_{\vec{k}{n}} =
\frac{2\pi}{\hbar} \int\sub{BZ} \frac{d\vec{k}'}{(2\pi)^3} \sum_{n'}
\left(f_{\vec{k}'n'}
    + n_T(\varepsilon_{\vec{k}'n'}-\varepsilon_{\vec{k}n}) \right)
\\ \times
\sum_{\vec{G}\vec{G}'}
\frac{1}{\pi}\Im\left[ \frac{4\pi e^2}{|\vec{k}'-\vec{k}+\vec{G}|^2}
	\epsilon^{-1}_{\vec{G}\vec{G}'}(\vec{k}'-\vec{k},\varepsilon_{\vec{k}n}-\varepsilon_{\vec{k}'n'}) \right]
\\ \times
\tilde{\rho}_{\vec{k}'n',\vec{k}n}(\vec{G})
\tilde{\rho}_{\vec{k}'n',\vec{k}n}^\ast(\vec{G}'),
\label{eqn:tauInv_ee}
\end{multline}
where $\tilde{\rho}_{\vec{q}'n',\vec{q}n}$ are density matrices in the plane-wave basis with reciprocal lattice vectors $\vec{G}$, and $\epsilon^{-1}_{\vec{G}\vec{G}'}$ is the inverse dielectric matrix calculated in the Random Phase Approximation.
See Ref.~\citenum{eeLinewidth} for a detailed introduction to this method and Ref.~\citenum{PhononAssisted} for our specific implementation details.
The only difference from Ref.~\citenum{PhononAssisted} is the factor $f_{\vec{k}'n'} + n_T(\varepsilon_{\vec{k}'n'}-\varepsilon_{\vec{k}n})$, where $n_T(\hbar\omega) \equiv \left(\exp\frac{\hbar\omega}{k_BT} - 1\right)^{-1}$ is the Bose function, which generalizes the approach from $T=0$ to the non-zero temperature $GW$ formalism.\cite{HighT-GW-SimpleMetal, FiniteT-GW}
In particular, this extension ensures a finite $\tau_{ee}^{-1} \propto \left((\varepsilon - \varepsilon_F)^2 + (\pi k_B T)^2\right)$ near the Fermi energy as expected for metals,\cite{TAparameters} while the conventional $T=0$ formalism for $\tau_{ee}$ diverges as $\varepsilon \to \varepsilon_F$.\cite{PhononAssisted}

The net carrier lifetime is $\tau^{-1}_{\vec{k}n} = (\tau\sub{e-ph}^{-1})_{\vec{k}n}
+ (\tau\sub{e-e}^{-1})_{\vec{k}n}$.
We use this to calculate the carrier linewidths
$\Im\Sigma_{\vec{k}n} = \hbar/(2\tau_{\vec{k}n})$ for the broadening of the energy conservation factors above, as well as for the mean free path $\lambda_{\vec{k}n} = |\vec{v}_{\vec{k}n}| \tau_{\vec{k}n}$.
Note that as in previous theoretical studies,\cite{brown2017experimental,NitrideCarriers} the carrier lifetime calculated here is the time for individual electron-phonon scattering events.
This is distinct from the electron-phonon fall/decay time often reported in pump-probe spectroscopy.\cite{su2019ultrafast,tomko2021temperature,hattori2021phonon} The electron-phonon relaxation time reported in experiments is the time constant for energy transfer from the electronic subsystem to the phononic subsystem.

\subsection{Interfacial injection probability}
\label{sec:MethodsInjection}

We estimate probability of injecting carriers into semiconductors using a modified Fowler model that accounts for the electronic structure of the metal by directly employing band momenta $\vec{p}_{\vec{k}n} = m_e\vec{v}_{\vec{k}n}$ computed from DFT (and interpolated using Wannier functions).\cite{GoldGaNcarriers, GoldGaNultrafast, CuGaNcarriers}
Specifically, for each electronic state in the metal, we decompose the momentum into a component along the interface normal direction (assumed to be the preferred transport direction $\hat{j}$), $p^\perp_{\vec{k}n} \equiv \vec{p}_{\vec{k}n} \cdot \hat{j}$, and a component parallel to the interface, $p^\parallel_{\vec{k}n} \equiv \left(|\vec{p}_{\vec{k}n}|^2 - (p^\perp_{\vec{k}n})^2\right)^{1/2}$.

The modified Fowler model matches the parallel component, requiring $p\sub{SC}^\parallel(\vec{p}_{\vec{k}n}) = p^\parallel_{\vec{k}n}$.
Assuming a parabolic band structure for the semiconductor, the corresponding interface-normal component of the momentum is
\begin{equation}
p_{SC}^\perp(\vec{p}_{\vec{k}n}, \varepsilon_{\vec{k}n})
= \begin{cases}
\sqrt{\Theta(2m^\ast_e(\varepsilon_{\vec{k}n} - \varepsilon_F - \phi_B^e) - (p^\parallel_{\vec{k}n})^2)},\\
\hfill	\varepsilon_{\vec{k}n} > \varepsilon_F + \phi_B^e \\
\sqrt{\Theta(2m^\ast_h(\varepsilon_F - \phi_B^h - \varepsilon_{\vec{k}n}) - (p^\parallel_{\vec{k}n})^2)},\\
\hfill	\varepsilon_{\vec{k}n} < \varepsilon_F - \phi_B^h \\
0, 
\hfill	\textrm{otherwise},
\end{cases}
\end{equation}
where $\Theta(x)$ is the Heaviside step function, $m^\ast_{e/h}$ and $\phi_B^{e/h}$ are the effective masses and Schottky barrier heights of electrons/holes in the semiconductor.
Note that states that are not allowed to enter the semiconductor lead to $p_{SC}^\perp = 0$.

Finally, the probability of injection in the modified Fowler model, accounting for a transmission matrix element due to momentum mismatch,\cite{HotCarrierReflection} is
\begin{equation}
P\super{inj}_{\vec{k}n} = \frac
{4 v^\perp_{\vec{k}n}
v\sub{SC}^\perp(\vec{v}_{\vec{k}n}, \varepsilon_{\vec{k}n})}
{\left(v^\perp_{\vec{k}n} + 
v\sub{SC}^\perp(\vec{v}_{\vec{k}n}, \varepsilon_{\vec{k}n})
\right)^2}.
\end{equation}
In our analysis in section~\ref{sec:Injection}, we report the average injection probability at each carrier energy,
\begin{equation}
P\sub{inj}(\varepsilon) = \frac{1}{g(\varepsilon)}
\int\sub{BZ} \frac{g\sub{s}\,\D\vec{k}}{(2\pi)^3} \sum_{n}
\delta(\varepsilon_{\vec{k}n} - \varepsilon) P\super{inj}_{\vec{k}n},
\end{equation}
where $g(\varepsilon)$ is the electronic density of states.
 \section{Results and Discussion}

\subsection{Directional conductors}

We recently identified several directional conductors that exhibit promise for conduction in nanoscale geometries using a high-throughput screening based on electronic structure calculations.\cite{DirectionalInterconnects}
These materials fall into two broad classes, film-like and wire-like conductors, which exhibit excellent transport in two perpendicular directions and along a single axis respectively.
To investigate the potential of directionality to additionally enhance harvesting of hot carriers far from equilibrium, we select two representative materials from each class.
In particular, for the film-like conductors, we select the delsafossite oxide \PtCoO with the lowest resistivity and the well-studied MAX phase \CrAlC, while for the wire-like conductors, we select the two most conductive materials, CoSn and \YCoB.

\begin{figure}
\centering
\includegraphics[width=\columnwidth]{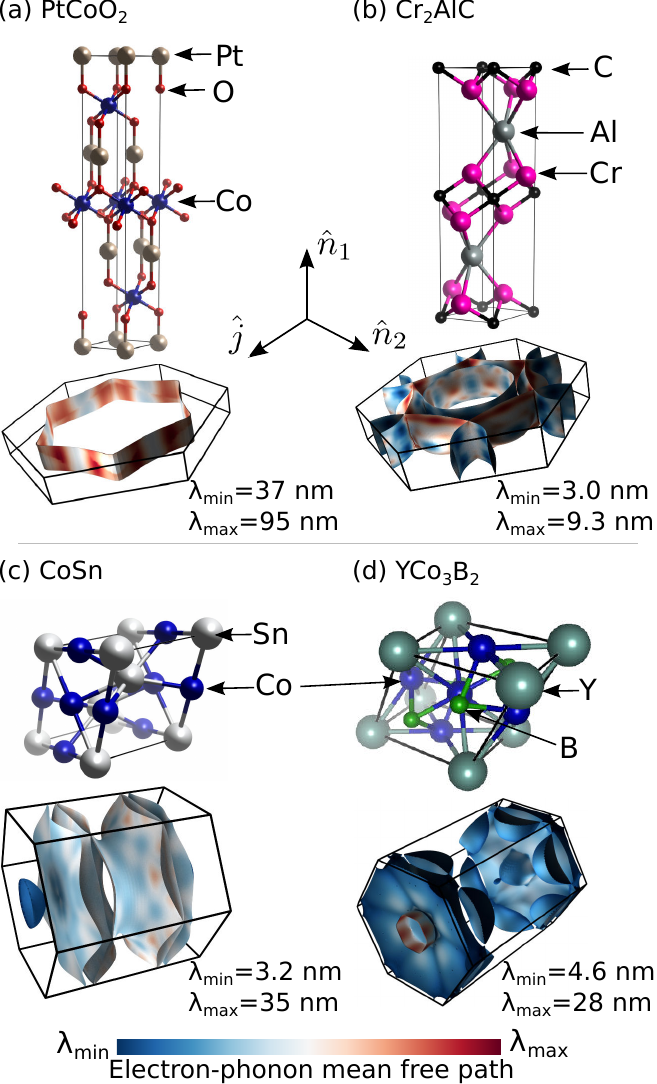}
\caption{Unit cell structure and Fermi surfaces of the directional conductors studied here.
Note that all are 3D hexagonal crystals, but have Fermi velocities (Fermi surface normals) along (a, b) the $a$ directions for the film-like conductors , and along (c, d) the $c$ direction for the wire-like conductors.
The electron-phonon mean free path $\lambda$ varies substantially over the Fermi surface for all four materials.}
\label{fig:FS} 
\end{figure}

Figure~\ref{fig:FS} shows the unit cell structure and Fermi surfaces for all four materials.
\PtCoO exhibits an almost perfect hexagonal-prism-shaped Fermi surface, which leads to all the Fermi velocities directed in the hexagonal plane of the material\cite{eyert2008metallic}.
This leads to an in-plane resistivity, along the preferred transport direction $\hat{j}$, that is $\sim 3,000\times$ smaller than the out-of-plane resistivity along the preferred surface normal direction $\hat{n}$ (Table~\ref{tab:resistivity}).
Such large anisotropies have been reported for transport properties of two-dimensional materials like graphene,\cite{pop2012thermal} attributed to the weak van der Waals coupling between layers.
However, all directional materials we investigate here are 3D crystals with strong bonding in all directions (Fig.~\ref{fig:FS}).
\CrAlC also has similar Fermi surface orientations and preferred transport directions, but a much weaker anisotropy in the resistivity\cite{bugnet2014experimental}.

\begin{table}
\centering
\begin{tabular}{cccccc}
\hline
\multirow{2}{*}{Type}
& \multirow{2}{*}{Material}
& \multirow{2}{*}{$\hat{j}$}
& \multirow{2}{*}{$\hat{n}$}
& \multicolumn{2}{c}{$\rho$ [$\mu\Omega\cdot$cm]}\\
\cline{5-6}
&&&& Along $\hat{j}$ & Along $\hat{n}$\\
\hline
\multirow{2}{*}{Film-like}
& \PtCoO & $\langle 1000\rangle$ & $[0001]$ & 1.8 & 5960 \\
& \CrAlC & $\langle 1000\rangle$ & $[0001]$ & 14.5 & 47.8 \\
\hline
\multirow{2}{*}{Wire-like}
& CoSn  & $[0001]$ & $\langle 1000\rangle$ & 2.9 & 37.2 \\
& \YCoB & $[0001]$ & $\langle 1000\rangle$ & 5.7 & 37.9 \\
\hline
\end{tabular}
\caption{Best transport direction $\hat{j}$, corresponding normal direction $\hat{n}$, and predicted resistivity along both directions, for the directional conductors investigated here.\cite{DirectionalInterconnects}
Note that the film-like and wire-like conductors are all 3D crystals with strong bonding in all directions (Fig.~\ref{fig:FS}), and not van der Waals / layered materials.}
\label{tab:resistivity}
\end{table}

The remaining two materials, CoSn and \YCoB, have Fermi surface sheets parallel to the hexagonal plane (Fig.~\ref{fig:FS}(c-d)), leading to Fermi velocities and conduction predominantly along the $c$ axis (Table~\ref{tab:resistivity}).
These materials therefore have a single preferred transport direction, with two preferred surface normal directions, making them suitable for conduction in nanoscale wires.\cite{DirectionalInterconnects}
To discuss these film-like and wire-like materials on a similar footing, we will henceforth refer to properties along preferred transport direction $\hat{j}$ and along the preferred normal direction $\hat{n}$, but note that the crystallographic orientation of these directions is different for the two classes of materials as specified in Table~\ref{tab:resistivity}.

Finally, note that the $\hat{j}$ resistivity for all these conductors, except for the \CrAlC max phase, is comparable to that for the most common plasmonic metals: 1.6, 1.7, 2.3 and 2.8 $\mu\Omega\cdot$cm for silver, copper, gold and aluminium,\cite{PhononAssisted} while the $\hat{n}$ resistivity is much larger.
This difference for Cr$_2$AlC can be understood by comparing the electronic structures of the four materials (see Fig.~S1 in supplemental material\cite{SI}.
In Cr$_2$AlC, the states near the Fermi level are dominated by $d$-bands~\cite{ito2017electronic} which tend to be narrower and hence, have smaller velocities and larger density of states at the Fermi level.
The other three materials studied here (PtCoO$_2$, YCo$_3$B$_2$ and CoSn) have larger Fermi velocities and smaller density of states due to larger dispersion of the bands crossing the Fermi level.
Consequently, Cr$_2$AlC exhibits higher electron-phonon scattering rates with more states to scatter into, leading to overall worse transport properties including bulk resistivity, carrier lifetimes and mean free paths.

\subsection{Optical response}

\begin{figure}[t]
\includegraphics[width=\columnwidth]{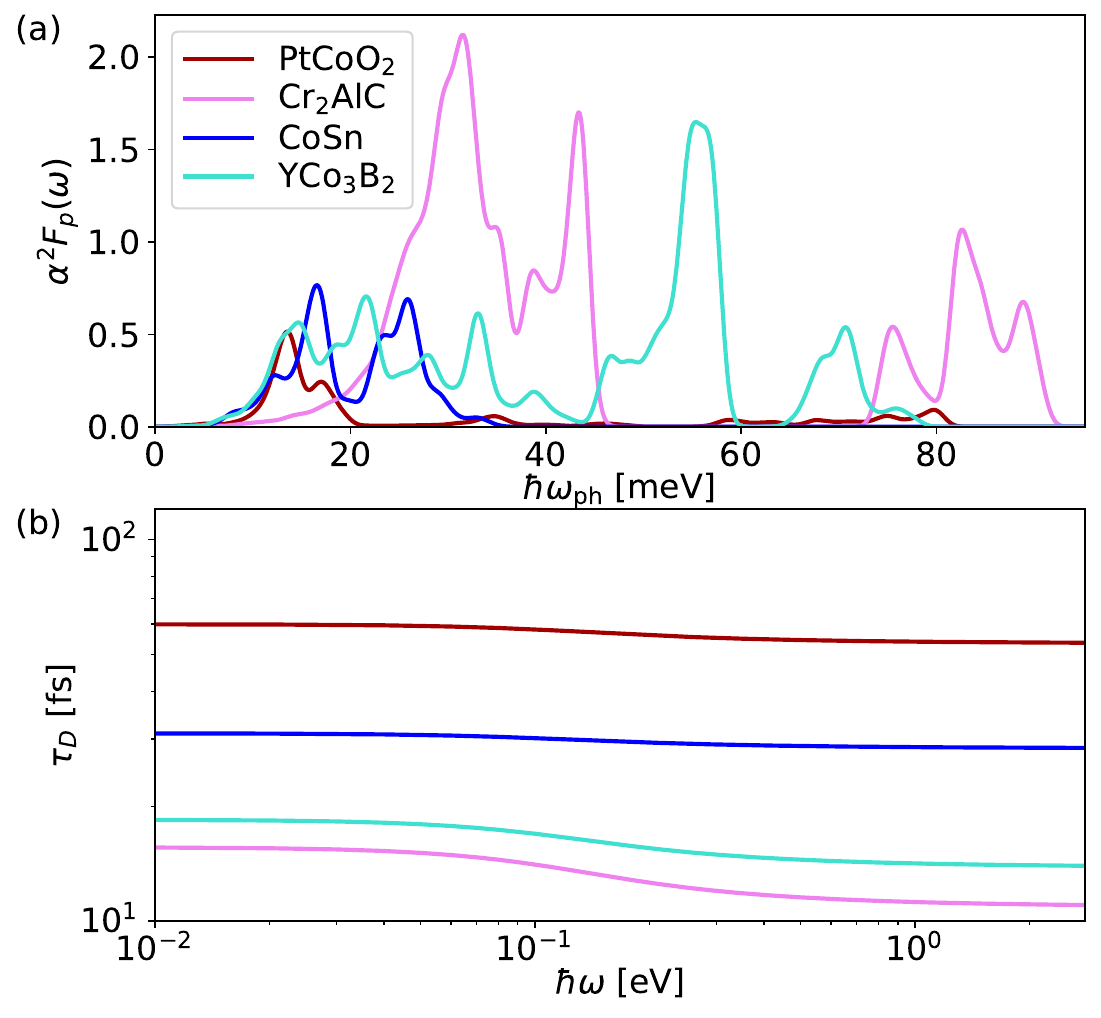}
\caption{(a) Transport-weighted Eliashberg spectral function and (b) frequency-dependent Drude relaxation time for directional conductors.
\PtCoO exhibits the lowest electron-phonon coupling strength (lowest $\alpha^2F(\omega)$) and hence the highest relaxation time $\tau_D$, while \CrAlC has the strongest coupling and lowest $\tau_D$.
The relaxation time drops from the $\omega = 0$ value for $\omega \sim \omega\sub{ph}$ and saturates to a high-frequency limit for $\omega \gg \omega\sub{ph}$ for each material.}
\label{fig:Ftau}
\end{figure}

To investigate the suitability of these materials for plasmonic applications, we first consider their Drude response.
Figure~\ref{fig:Ftau} shows the calculated transport Eliashberg spectral functions (Section~\ref{sec:MethodsDielectric}) and corresponding frequency-dependent Drude momentum relaxation times due to electron-phonon scattering.
The electron-phonon coupling at the Fermi level is weakest for \PtCoO, as indicated by the lowest spectral function magnitude in Fig.~\ref{fig:Ftau}(a), followed by CoSn, \YCoB and finally \CrAlC.
Correspondingly, the Drude relaxation time $\tau_D$ at low frequencies drops from $\sim 60$~fs for \PtCoO to $\sim 18$~fs for \YCoB, in the same order.

With increasing frequency, Drude relaxation can access higher energy phonon modes that were inaccessible using thermal energy at room temperature alone, causing $\tau_D$ to decrease.
For \PtCoO, $\tau_D$ decreases from 60~fs at low frequency to $\sim$54~fs at optical frequencies, remaining larger than the low-frequency relaxation time for most elemental metals!
The largest proportional drop in $\tau_D$ is from 15.6~fs to 11~fs for \CrAlC, which has the highest-energy phonon spectrum, while the smallest drop in $\tau_D$ is from 31~fs to 28~fs for CoSn with the lowest-energy phonon spectrum.

\begin{figure}
\includegraphics[width=\columnwidth]{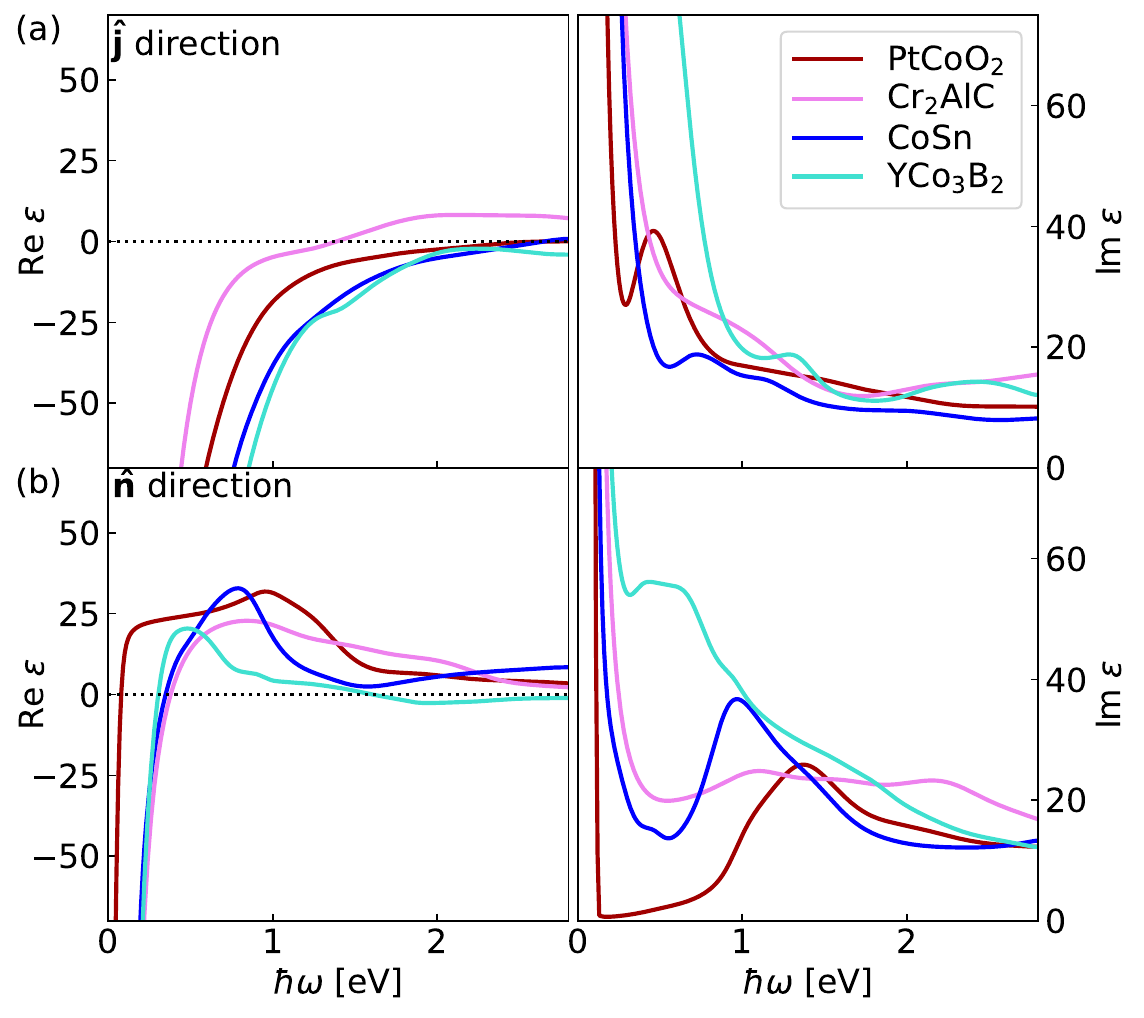}
\caption{Complex dielectric function for the directional conductors along (a) the preferred transport direction $\hat{j}$ and (b) the corresponding normal direction $\hat{n}$, with real and imaginary parts in the left and right columns respectively.
The plasmonic window ($\Re\epsilon < 0$) is much wider along the $\hat{j}$ direction than the $\hat{n}$ direction in all cases, making the optical response of the film conductors \PtCoO and \CrAlC similar to 2D materials and that of wire conductors CoSn and \YCoB similar to 1D materials, even though all four materials are 3D materials with strong bonding in all directions.}
\label{fig:Epsilon} 
\end{figure}

The decrease in $\tau_D$ with frequency accounts for phonon-assisted transitions along with the Drude term,\cite{Argentene} which we then combine with predictions of direct transitions (Section~\ref{sec:MethodsDielectric}) to compute the overall complex dielectric functions shown in Fig.~\ref{fig:Epsilon}.
Note that the dielectric functions are highly anisotropic, with qualitative differences between the $\hat{j}$ and $\hat{n}$ directions for these directional conductors.
In particular, the plasmonic window -- the region with $\Re\epsilon < 0$ is much larger along the $\hat{j}$ direction than the $\hat{n}$ direction.
The plasmonic window of \PtCoO, CoSn, and \YCoB each extend above 2.5 eV along $\hat{j}$, surpassing the plasmonic range of gold (2.6 eV) and approaching that of silver (3.4 eV).\cite{TAparameters}
In contrast, the plasmonic window along $\hat{n}$ is below 0.5 eV for all four materials, with that of \PtCoO (the most anisotropic of the four) being virtually negligible.

\begin{figure}[t]
\includegraphics[width=\columnwidth]{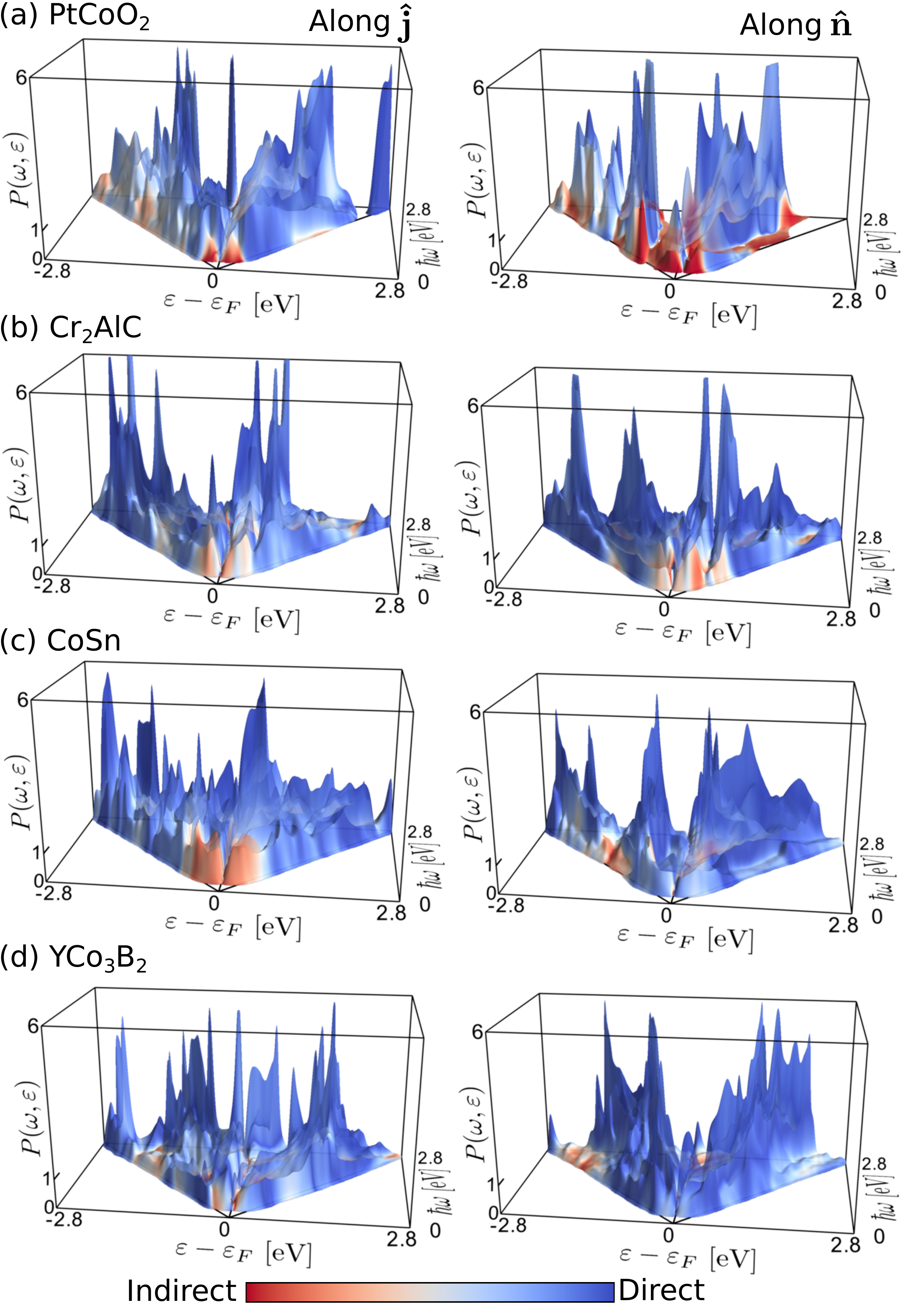}
\caption{Hot carrier distribution, $P(\omega, \varepsilon)$, as a function of carrier energy $\varepsilon$ and plasmon frequency $\omega$ for directional conductors, with plasmon electric field polarized along their corresponding $\hat{j}$ directions (left panels) and $\hat{n}$ directions (right panels).
The probability is normalized such that $P(\omega, \varepsilon) = 1$for a uniform electron and hole energy distribution.\cite{PhononAssisted}
Note that direct transitions dominate at almost all plasmon frequencies, except for $\hbar\omega\ll 1$~eV in \PtCoO and CoSn, leading to carrier distributions sensitive to electronic structure details.}
\label{fig:carrierDistribution} 
\end{figure}

Additionally, $\Im\epsilon$ along the preferred transport directions are relatively low when Re $\epsilon$ crosses zero, indicating plasmonic response with quality factors comparable to noble metals, most notably for CoSn.
In the plasmonic frequency range for $\hat{j}$, most of these materials exhibit dielectric-like behavior, $\Re\epsilon \gg \Im\epsilon > 0$, along the $\hat{n}$ direction.
Consequently, these materials essentially behave like lower-dimensional materials optically -- the film-like conductors like 2D materials and the wire-like conductors like 1D materials, but we emphasize again that these materials are 3D crystals bonded strongly in all directions.
This opens up the possibility to reap the advantages of low-D plasmonics, such as high confinement,\cite{Argentene} while working with easier-to-process 3D materials.
See the supplemental material for tables of predicted dielectric functions to facilitate design of plasmonic structures using these materials;\cite{SI} we focus here instead on material properties including hot carrier distributions and transport next.

\subsection{Hot carrier generation}

Plasmon decay is captured by the imaginary part of the dielectric functions, $\Im\epsilon(\omega)$, and the energy $\hbar\omega$ lost by the plasmon is deposited in an electron-hole pair.
Figure~\ref{fig:carrierDistribution} shows the energy distribution of hot carriers excited upon plasmon decay, calculated by resolving contributions to $\Im\epsilon(\omega)$ by carrier energy $\varepsilon$ (Section~\ref{sec:MethodsDielectric}).\cite{PhononAssisted}
Unlike a clear separation between phonon-assisted transitions below an interband threshold and direct transition above it in typical plasmonic metals,\cite{PhononAssisted, NitrideCarriers} these complex conductors exhibit direct transitions at almost all plasmon frequencies.
Only \PtCoO and CoSn show a small frequency range ($\hbar\omega < 0.5$~eV) where indirect phonon-assisted transitions contribute significantly.
Interestingly, the relative importance of phonon-assisted transitions in \PtCoO is enhanced along the $\hat{n}$ direction, indicating a selection rule that the lowest energy direct transitions are allowed only for in-plane electric fields in this material.

\begin{figure}
\includegraphics[width=\columnwidth]{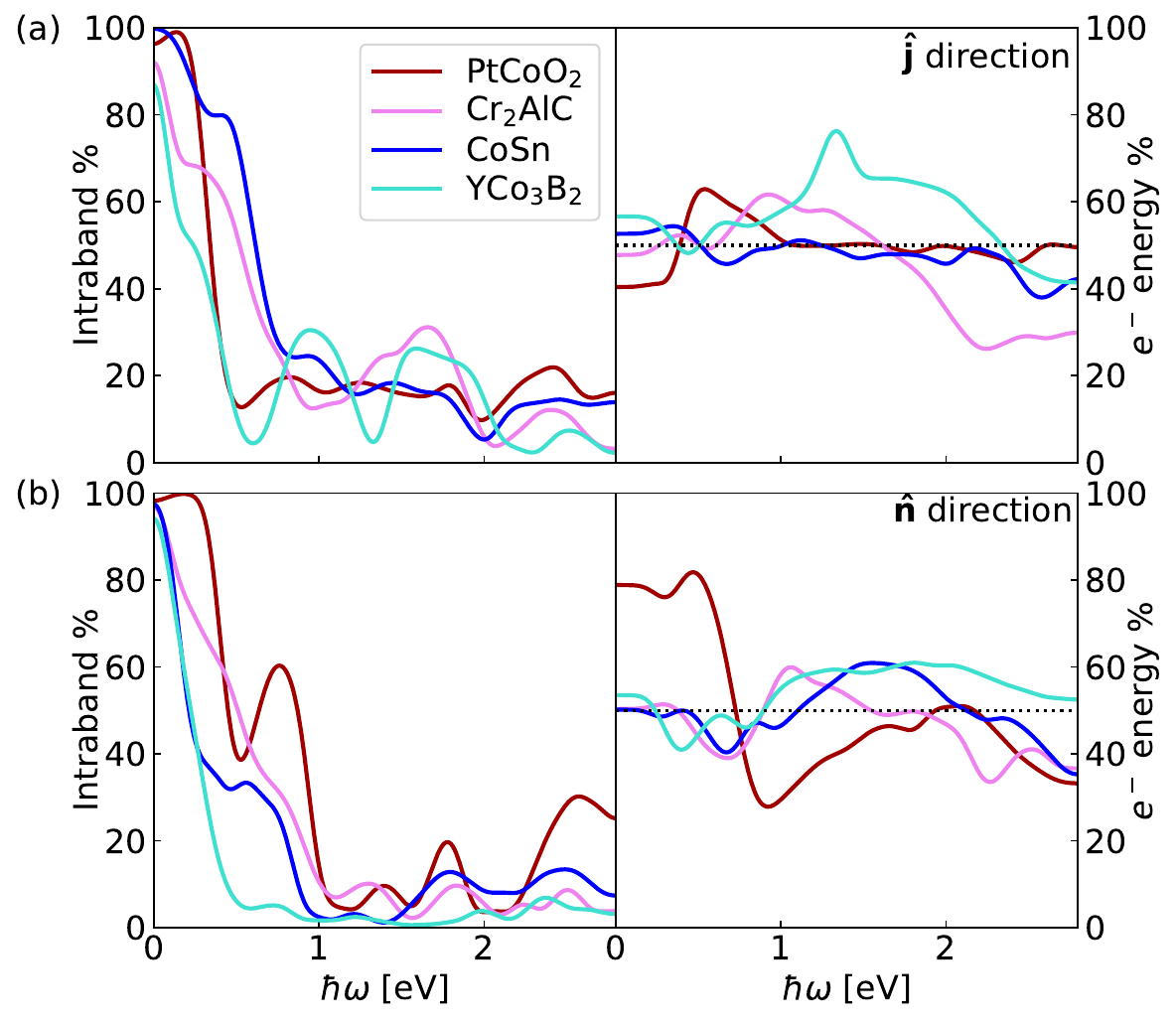}
\caption{Fraction of interband absorption (left column) and fraction of energy in hot electrons (right column) for the directional conductors for light polarized along (a) the preferred transport direction $\hat{j}$ and (b) the corresponding normal direction $\hat{n}$.
Interband absorption is stronger along the $\hat{j}$ direction, but is dominated by direct absorption for photons with energy larger than 1~eV in all cases.
Energy is evenly distributed between electrons and holes, except for low-energy intraband-dominated absorption of $\hat{n}$-polarized light in \PtCoO.}
\label{fig:carrierStats}
\end{figure}

The predominance of direct transitions leads to relatively complex energy distributions of carriers, sensitive to details in the electronic structure.
Consequently, Figure~\ref{fig:carrierStats} shows moments of these carrier distributions, indicating the fraction of (indirect) intraband transitions and the distribution of energy between electrons and holes.
Direct transitions dominate above 1~eV, with the intraband fraction dropping below 20\%, in all four materials.
Additionally, despite the complex and different structure in the electron and hole distributions in Fig.~\ref{fig:carrierDistribution}, the overall energy separation between electrons and holes is approximately even, with between 40 and 60\% of the energy in the electrons, across the frequency range considered for all materials.
The only notable deviation is that 80\% of the energy is deposited in the electrons by intraband transitions in \PtCoO for $\hat{n}$ polarization, while the energy split is close to equal even in that energy range for the $\hat{j}$ polarization.
Overall, given the absence of systematic asymmetries in electron and hole distributions, such as those for noble metals,\cite{PhononAssisted} analyses of hot carriers in these directional conductors may safely assume a uniform electron and hole energy distribution extending from 0 to the plasmon energy, independent of polarization.

Note that the hot carrier properties presented here are calculated for bulk materials.
Since surface sites (like edges, facets and corners) and shape and size of the nanoparticles can significantly enhance the rate and distribution of hot carrier generation,\cite{rossi2020hot} future studies should focus on the effect of atomic structure on plasmonic properties of these directional conductors.
Additionally, nanoparticle shapes and surface characteristics will vary dramatically depending on the synthesis approach, and present additional opportunities in targeting techniques such as self-organized nanodecomposition\cite{SelfOrganized} towards nanoparticles suitable for hot carrier extraction in future work.

\subsection{Carrier transport}

\begin{figure}
\includegraphics[width=\columnwidth]{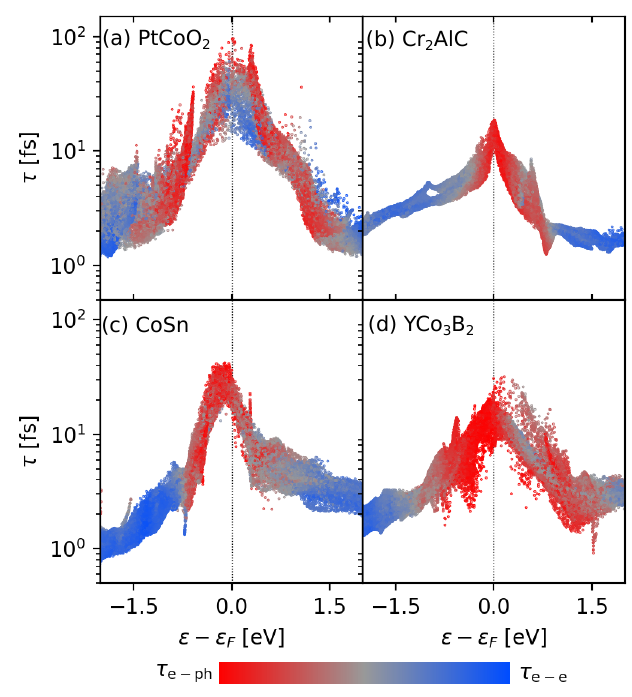}
\caption{Lifetime of hot carriers as a function of carrier energy in directional conductors, with color indicating relative contributions of electron-phonon and electron-electron scattering.
Carrier lifetimes are comparable in magnitude to noble metals, especially in \PtCoO and CoSn, but drop off with energy faster than for noble metals.\cite{PhononAssisted}}
\label{fig:tau}
\end{figure}

\begin{figure}
\includegraphics[width=\columnwidth]{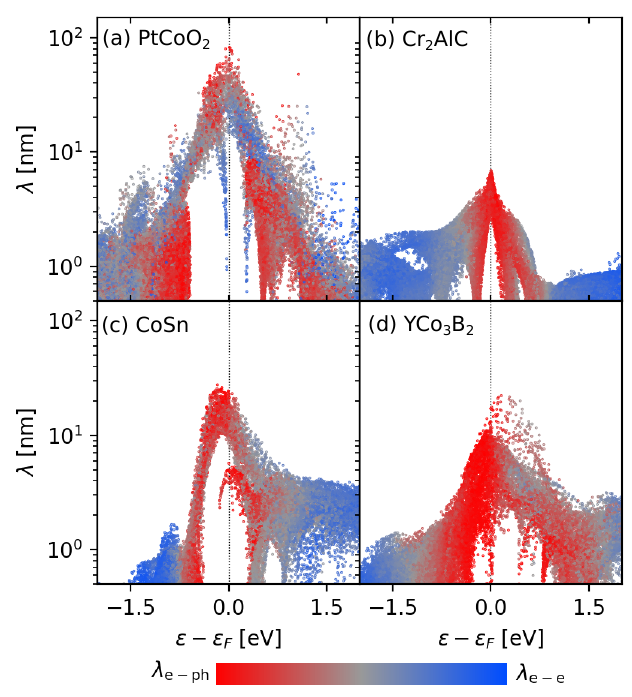}
\caption{Mean-free path of hot carriers as a function of carrier energy in directional conductors, with behavior closely following those of the lifetimes shown in Fig.~\ref{fig:tau}.}
\label{fig:lambda}
\end{figure}

Efficient hot carrier harvesting requires extraction of the carriers from the metal prior to thermalization by electron-phonon and electron-electron scattering.
Figures~\ref{fig:tau} and \ref{fig:lambda} respectively show the predicted lifetime and mean free path of hot carriers in the directional conductors, accounting for both these scattering mechanisms (Section~\ref{sec:MethodsLifetime}).
Electron-phonon scattering dominates near the Fermi energy, while the phase-space for electron-electron scattering increases with increasing electron or hole energies and eventually dominates over electron-phonon scattering, exactly analogous to the case for other plasmonic metals including elemental metals and transition metal nitrides.\cite{PhononAssisted, NitrideCarriers}
The mean free paths drop off with energy more quickly than the life times in most of the materials, especially for the holes, indicating a reduction in average carrier velocities for energies far from the Fermi level.

The peak carrier lifetimes and mean free paths in these materials are competitive with noble metals, especially in \PtCoO and CoSn.
However, the drop-off with energy is faster in these materials compared to noble metals, with mean-free paths dropping below 10~nm for carriers with 1~eV energies for all all four materials.
This is partly because the directionality of electronic structure is a feature restricted to a narrow energy window surrounding the Fermi energy, with a marked increase in density of states away from the Fermi energy (Figure~S1 in Supplemental Material\cite{SI}), leading to a corresponding increase in pase space for scattering for higher energy carriers.
This indicates the need to carefully design plasmonic geometries to minimize the average distance between the location of carrier generation and the interfaces where carriers can be collected to below 10~nm.

\subsection{Carrier injection across interfaces}
\label{sec:Injection}

\begin{figure}[t]
\includegraphics[width=\columnwidth]{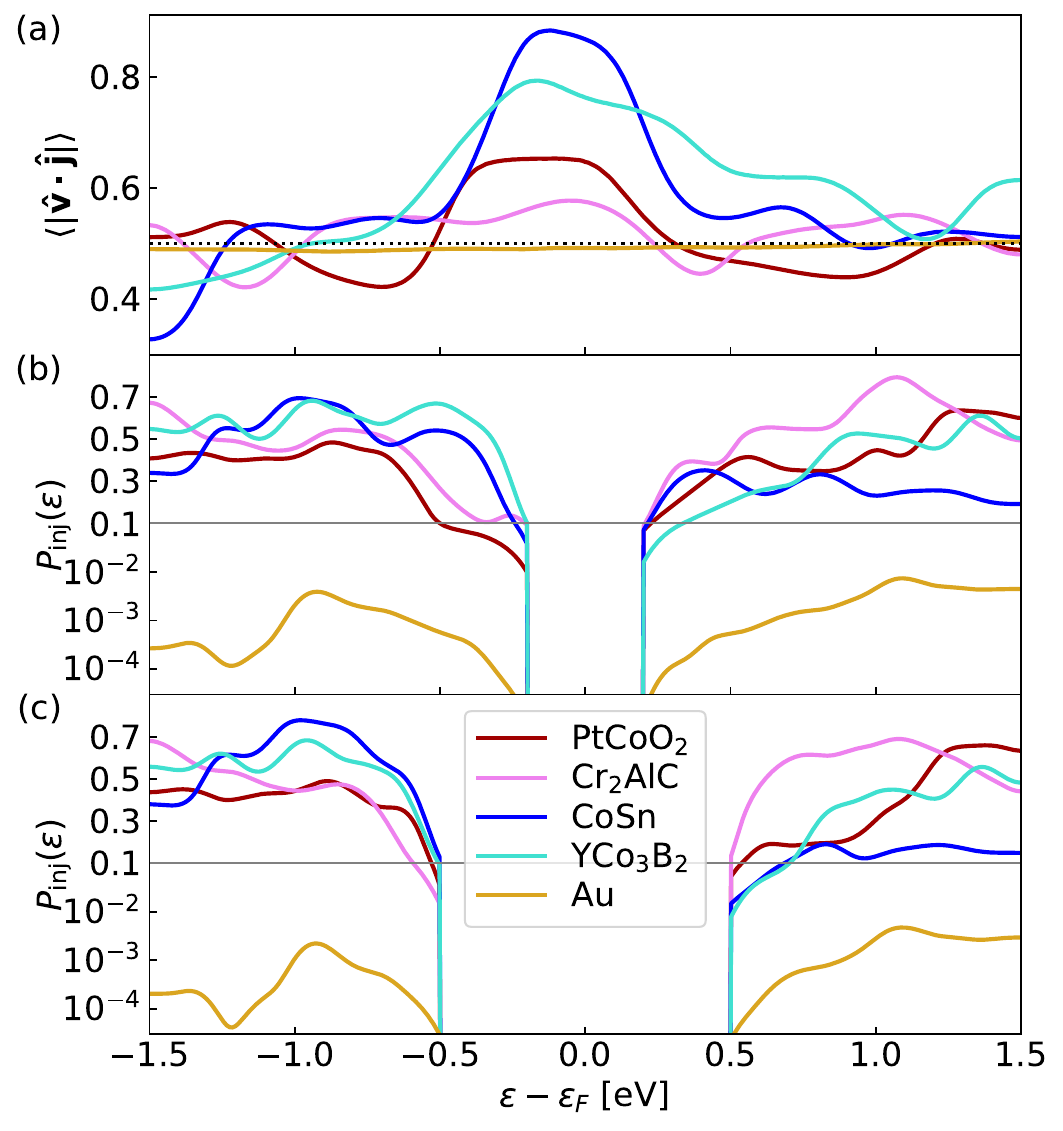}
\caption{(a) Average velocity direction $\hat{v}$ component along preferred transport direction $\hat{j}$ in directional conductors compared to gold, as a function of carrier energy.
(The dotted line at $\langle|\hat{v}\cdot\hat{j}|\rangle = 1/2$ indicates the isotropic limit.)
The preferred orientation of velocities substantially increases the probability of carrier injection into a semiconductor (assumed parabolic bands with $m^\ast = 0.3 m_e$) with Schottky barrier heights (b) 0.2~eV and (c) 0.5~eV.
Note the switch from logarithmic to linear scaling in the $y$-axis of (b,c) to simultaneously show the small injection probabilities from gold and the much larger injection probabilities from all directional conductors.}
\label{fig:injection}
\end{figure}

The final step in hot carrier harvesting is transferring the carriers across an interface, such as a metal-semiconductor interface, prior to energy relaxation in the metal.
This step is typically the largest limiting factor in hot carrier collection efficiency because most carriers in the metal encounter total internal reflection at the interface with the semiconductor, as captured by the semi-classical Fowler model of carrier injection.\cite{HotCarrierReflection, ScratchTheSurface}
In particular, the injection from noble metals into semiconductors at energies just above the Schottky barriers is typically $\sim 0.1$\% or smaller due to the large mismatch between the Fermi velocity in the metal and velocities near the band edges of the semiconductor.\cite{GoldGaNcarriers}
Injection from lower-velocity bands of the metal, such as $d$-bands in copper, can slightly increase these probabilities, but they remain at that small order of magnitude.\cite{CuGaNcarriers}

The directional conductors investigated here have the potential to remedy this issue in two ways.
First, the Fermi velocity in these materials is lower than in noble metals, exhibiting a slightly better match to velocities near semiconductor band edges.
Second, the directionality of the velocity makes it possible to substantially reduce the component parallel to the surface (that needs to match across the interface) by collecting in an interface whose normal is along $\hat{j}$.

Figure~\ref{fig:injection}(a) shows that the component of carrier velocity direction $\hat{v}$ along $\hat{j}$ is close to 1 for wire-like conductors CoSn and \YCoB, indicating that the remaining components must be much smaller.
This projection is smaller for the film-like conductors, which have large velocity components along two orthogonal directions, and approaches the isotropic limit of 1/2 for a noble metal like gold.
Figures~\ref{fig:injection}(b-c) show that this directionality of the velocity indeed leads to a significantly larger injection probability than for the case of gold, as predicted using a modified Fowler model (Section~\ref{sec:MethodsInjection}).
Compared to $P\sub{inj}\sim 10^{-3}$ in gold, all directional conductors have the potential to inject with $> 10\%$ probability into a model semiconductor (assuming a typical electron and hole mass of 0.3 for simplicity), presenting significant opportunities for efficient plasmonic hot carrier harvesting.
 \section{Conclusions} \label{sec:conclusions}
Using first-principles calculations of optical response, carrier generation, transport and injection, we show the significant promise of directional conductors for plasmonic hot carrier applications.
The strong anisotropy of Fermi velocity in these materials to one or two directions introduces 1D wire-like or 2D film-like behavior in a 3D-crystalline material.
These materials exhibit plasmonic response over a range of frequencies comparable to noble metals along the preferred transport direction(s) $\hat{j}$, while simultaneously exhibiting dielectric behavior in the preferred normal direction(s) $\hat{n}$ at those frequencies, mimicking the optical response of a lower-dimensional metal.
Hot carrier generation and transport are also similar to noble metals, but directionality in the momenta shows promise for orders of magnitude higher efficiency in extraction of hot carriers from these materials into semiconductors.
The dielectric functions, carrier transport and injection models reported here can now be used to design plasmonic nanostructures with the potential for high-efficiency hot carrier harvesting.  \section*{Acknowledgements}
The authors acknowledge funding from SRC under Task No. 2966.002.
Calculations were carried out at the Center for Computational Innovations at Rensselaer Polytechnic Institute.
 
 \end{document}